\documentclass[a4paper,fleqn]{article}
\pdfoutput=1
\usepackage{amsmath}
\usepackage[hidelinks]{hyperref}

\begin{document}

\title{\textbf{On a new 3D generalized Hunter--Saxton equation}}

\author{\textsc{Sergei Sakovich}\bigskip \\
\small Institute of Physics, National Academy of Sciences of Belarus \\
\small \href{mailto:sergsako@gmail.com}{sergsako@gmail.com}}

\date{}

\maketitle

\begin{abstract}
The problem of integrability is studied for a 3D generalized Hunter--Saxton equation introduced recently by O.I.~Morozov. A transformation is found which brings the equation into a constant-characteristic form and simultaneously trivializes the equation's Lax representation. The transformed equation is shown to fail the Painlev\'{e} test for integrability.
\end{abstract}

\section{Introduction} \label{s1}

In this paper, we study the three-dimensional nonlinear wave equation
\begin{equation}
u_{xt} = u u_{xx} - u_x^2 + u_y \label{e1}
\end{equation}
introduced recently in \cite{M} and called the 3D generalized Hunter--Saxton equation there. This equation is considered as integrable in \cite{M} because it is associated with an infinite-dimensional Lie algebra which generates the over-determined linear problem
\begin{equation}
r_t = u r_x , \qquad r_y = u_x r_x \label{e2}
\end{equation}
called the Lax representation of \eqref{e1}. Note that we have corrected the sign of the last term in \eqref{e1} because the equation with $- u_y$ as given in \cite{M} would not match the linear problem \eqref{e2}.

The reduction of \eqref{e1} with $u_y = 0$ is the equation $u_{xt} = u u_{xx} - u_x^2$ (not the Hunter--Saxton equation \cite{HS} $u_{xt} = u u_{xx} + \frac{1}{2} u_x^2$) which belongs to the Calogero's class of exactly solvable by quadratures equations \cite{C} $u_{xt} = u u_{xx} + F( u_x )$ with any function $F$. The existence of this exactly solvable two-dimensional reduction is an interesting feature but it tells nothing about the integrability of the three-dimensional equation \eqref{e1} itself. It also remains unknown what the parameterless first-order scalar linear problem \eqref{e2} means and how it can be used to integrate the nonlinear equation \eqref{e1}. In our experience, statements about the integrability of nonlinear equations associated with infinite-dimensional Lie algebras should be considered with caution \cite{S17,S21}.  Definitely, the integrability of \eqref{e1} deserves further investigation.

The paper is organized as follows. In Section~\ref{s2}, we find a transformation which relates the nonlinear equation \eqref{e1} with a simpler equation possessing constant characteristics and simultaneously makes the Lax representation \eqref{e2} trivial. In Section~\ref{s3}, we carry out the singularity analysis of the nonlinear equation obtained via the transformation and show that the equation fails the Painlev\'{e} test for integrability. Section~\ref{s4} contains concluding remarks.

\section{Transformation} \label{s2}

To transform the nonlinear equation \eqref{e1} into a simpler equation with constant characteristics, we follow the way successfully used in \cite{SS07,S09,S16,S18a,S18b} for a series of other nonlinear equations. However, now we do it for the first time with a three-dimensional equation.

The change of variables
\begin{equation}
x = w(z,y,t) , \qquad u(w(z,y,t),y,t) = v(z,y,t) , \label{e3}
\end{equation}
made in \eqref{e1}, leads us to
\begin{gather}
v_{zt} - \frac{w_t + v}{w_z} v_{zz} + \frac{v_z^2}{w_z} - w_z v_y \notag \\
+ \left( \frac{w_t + v}{w_z^2} w_{zz} - \frac{w_{zt}}{w_z} + w_y \right) v_z = 0 , \label{e4}
\end{gather}
where $w(z,y,t)$ is arbitrary (with $w_z \neq 0$, of course). We see that the choice of $w : \, w_t = -v$ would eliminate the term with $v_{zz}$ and simplify \eqref{e4} as
\begin{equation}
v_{zt} + 2 \frac{v_z^2}{w_z} - w_z v_y + w_y v_z = 0 . \label{e5}
\end{equation}
Therefore we make the substitution
\begin{equation}
v = - w_t \label{e6}
\end{equation}
in \eqref{e4} (or, equivalently, in \eqref{e5}), multiply the result by $- 1 / w_z^2$, and get
\begin{equation}
\partial_t \left( \frac{w_{zt}}{w_z^2} - \frac{w_y}{w_z} \right) = 0 , \label{e7}
\end{equation}
which is integrated as
\begin{equation}
\frac{w_{zt}}{w_z^2} - \frac{w_y}{w_z} = h(z,y) , \label{e8}
\end{equation}
where $h(z,y)$ is arbitrary.

Without loss of generality, we can choose $h(z,y) = 0$ in \eqref{e8}, for the following reason. The relations \eqref{e3}, $x = w(z,y,t)$ and $u = v(z,y,t)$, can be considered as a parametric representation of solutions $u(x,y,t)$ of \eqref{e1}, where $z$ serves as the parameter and the arbitrariness of $w(z,y,t)$ corresponds to the arbitrariness of the parameter's choice, $z \mapsto f(z,y,t)$ with arbitrary $f(z,y,t)$. When we choose $w(z,y,t)$ to satisfy \eqref{e6}, there still remains the arbitrariness $z \mapsto g(z,y)$ of the parameter's choice, with arbitrary $g(z,y)$. This change of $z$, $z \mapsto g(z,y)$, generates the change of $h(z,y)$ in \eqref{e8}, $h(z,y) \mapsto g_y + h(z,y) g_z$, therefore we can always make $h = 0$ in \eqref{e8} by an appropriate choice of the parameter $z$. Moreover, even when we fix $h = 0$ in \eqref{e8}, there still remains the arbitrariness $z \mapsto a(z)$ of the parameter's choice, with arbitrary $a(z)$.

Consequently, all solutions $u(x,y,t)$ of the nonlinear equation \eqref{e1} are parametrically represented by all solutions $w(z,y,t)$ (with $w_z \neq 0$) of the nonlinear equation
\begin{equation}
w_{zt} = w_z w_y \label{e9}
\end{equation}
via the relations
\begin{equation}
x = w(z,y,t) , \qquad u(x,y,t) = - \partial_t w(z,y,t) , \label{e10}
\end{equation}
where $z$ serves as the parameter. The arbitrariness $z \mapsto a(z)$ of the parameter's choice, with any $a(z)$, corresponds to the invariance of \eqref{e9} and has no effect on solutions of \eqref{e1}. Of course, all our consideration was purely local.

It is interesting to see how the transformation \eqref{e10}, relating \eqref{e1} with \eqref{e9}, acts on the (so-called) Lax representation \eqref{e2}. We introduce the function $s(z,y,t)$, such that
\begin{equation}
r(w(z,y,t),y,t) = s(z,y,t) , \label{e11}
\end{equation}
and obtain from \eqref{e2},  \eqref{e10} and  \eqref{e9} the trivial linear system
\begin{equation}
s_t = 0 , \qquad s_y = 0 . \label{e12}
\end{equation}
Since $s=s(z)$ follows from \eqref{e12}, the linear system \eqref{e2} may be somehow related to the transformation \eqref{e10} we found, but it definitely tells nothing about how to integrate the nonlinear equations \eqref{e1} and \eqref{e9}.

\section{Singularity analysis} \label{s3}

We have never seen the nonlinear equation \eqref{e9} in the literature. Let us study its integrability by the Painlev\'{e} test for partial differential equations \cite{WTC,T}. In our experience, based on the singularity analysis of wide classes of nonlinear systems \cite{S99,ST,KKS01,KKSST}, the Painlev\'{e} test is a reliable and convenient tool, capable not only to detect all known integrable cases but also to discover some interesting new ones \cite{KKS04,K08,S19}.

A hypersurface $\phi(z,y,t) = 0$ is non-characteristic for \eqref{e9} if $\phi_z \phi_t \neq 0$. Near a non-characteristic hypersurface $\phi = 0$, the dominant singular behavior of solutions $w$ of the nonlinear equation \eqref{e9} is
\begin{equation}
w = - \frac{\phi_t}{\phi_y} \log \phi + \dotsb . \label{e13}
\end{equation}
Note that \eqref{e13} does not work if $\phi_y = 0$ (perhaps because \eqref{e9} with $w_y = 0$ is the linear equation $w_{zt} = 0$ whose solutions have singularities at the characteristics only). In what follows, we consider the generic case with $\phi_y \neq 0$, and we set $\phi_y = 1$ without loss of generality,
\begin{equation}
\phi = y + \psi(z,t) , \qquad \psi_z \psi_t \neq 0 . \label{e14}
\end{equation}

This dominant logarithmic singularity \eqref{e13} is not a good starting point for the Painlev\'{e} analysis. The situation, however, can be improved by the new dependent variable $q(z,y,t)$,
\begin{equation}
q = w_z , \label{e15}
\end{equation}
which turns the nonlinear equation \eqref{e9} into
\begin{equation}
q q_{zt} - q_z q_t -q^2 q_y = 0 . \label{e16}
\end{equation}
Near a hypersurface $\phi = 0$ with $\phi$ given by \eqref{e14}, the dominant singular behavior of solutions $q$ of the nonlinear equation \eqref{e16} is
\begin{equation}
q = - \psi_z \psi_t \, \phi^{-1} + \dotsb . \label{e17}
\end{equation}

Now we can try to represent the general solution of \eqref{e16} near $\phi = 0$ by the generalized Laurent series
\begin{equation}
q = q_0 (z,t) \phi^{-1} + \dotsb + q_i (z,t) \phi^{i-1} + \dotsb , \label{e18}
\end{equation}
where $\phi$ is given by \eqref{e14}. We substitute the expansion \eqref{e18} to the nonlinear equation \eqref{e16}, collect terms with $\phi^{n-4}$, separately for $n = 0,1,2, \dotsc$, and obtain in this way the following. The resonances, where arbitrary functions can enter the expansion \eqref{e18}, turn out to be
\begin{equation}
n = -1 , 1 , \label{e19}
\end{equation}
and $n = -1$, as always, corresponds to the arbitrariness of $\psi(z,t)$ in $\phi$ \eqref{e14}. At $n = 0$, we get
\begin{equation}
q_0 = - \psi_z \psi_t , \label{e20}
\end{equation}
as expected due to \eqref{e17}. However, at $n = 1$, where we have the resonance and the function $q_1 (z,t)$ remains undetermined (arbitrary), we get the nontrivial compatibility condition
\begin{equation}
\psi_{zt} = 0 , \label{e21}
\end{equation}
which shows that the expansion \eqref{e18} is valid only for a quite restricted class of hypersurfaces $\phi = 0$. Consequently, the nonlinear equation \eqref{e16} has failed the Painlev\'{e} test for integrability.

It is interesting to see what is the valid expansion for the general solution of the nonlinear equation \eqref{e16}. To avoid the appearance of the nontrivial compatibility condition \eqref{e21}, we have to modify the expansion \eqref{e18} by a logarithmic term introduced before the resonance term,
\begin{equation}
q = q_0 (z,t) \phi^{-1} + b(z,t) \log \phi + q_1 (z,t) + \dotsb . \label{e22}
\end{equation}
It follows from \eqref{e16} and \eqref{e22} that $q_0$ is given by \eqref{e20},
\begin{equation}
b = - \frac{1}{2} \psi_{zt} , \label{e23}
\end{equation}
and $q_1 (z,t)$ remains arbitrary. All the higher-order terms of the expansion \eqref{e22} are determined by \eqref{e16} recursively, in terms of two arbitrary functions, $\psi (z,t)$ and $q_1 (z,t)$, and their derivatives. For example, the next three terms of the expansion \eqref{e22} are
\begin{equation}
c(z,t) ( \log \phi )^2 \phi + d(z,t) ( \log \phi ) \phi + q_2 (z,t) \phi , \label{e24}
\end{equation}
where
\begin{equation}
c = - \frac{\psi_{zt}^2}{12 \psi_z \psi_t} , \label{e25}
\end{equation}
but we omit the expressions for $d$ and $q_2$ as cumbersome and unnecessary. We have got a so-called logarithmic psi-series \cite{T}. Such expansions are typical for nonlinear equation considered as non-integrable (at least, currently).

\section{Conclusion} \label{s4}

Summarizing the obtained results, we can state that the 3D generalized Hunter--Saxton equation is most probably non-integrable. We also believe that it may be more convenient and productive to further investigate this equation in its equivalent form $w_{zt} = w_z w_y$ we found.

\end{document}